\documentclass[12pt]{article}
\usepackage{subeqnarray,eqsection,indent,cite}
\usepackage{latexsym}
\usepackage{amsmath}           % essential
\usepackage{amsfonts}          % AMS fonts
\usepackage{amssymb}           % AMS symbols
\usepackage{mathrsfs}

\newcommand{\eMut}{e^{M_{t+1} }U_{0,t}^\dag}
\newcommand{\eMvt}{U_{0,t} e^{M_{t+1}}}
\newcommand{\emenMut}{U_{0,t} e^{-M_{t+1}} }
\newcommand{\emenMvt}{ e^{-M_{t+1}}U_{0,t}^\dag}
\newcommand{\eMvtmenone}{U_{0,t-1} e^{M_{t}}}
\newcommand{\meta}{\frac{ 1}{2}\,}
\newcommand{\uh}{{\hat u}}
\newcommand{\vh}{{\hat v}}

\newcommand{\bone}{\mbox{$1 \hspace{-1.0mm} {\bf l}$}}

\newcommand{\alphah}{{\hat \alpha}}
\newcommand{\betah}{{\hat \beta}}

\newcommand{\tr}{\mathop{\rm tr}\nolimits}
\newcommand{\Tr}{\mathop{\rm Tr}\nolimits}

\newcommand{\cof}{\mathop{\rm cof}\nolimits}
\newcommand{\suf}{\mathop{\rm sf}\nolimits}

\newcommand{\fc}{{\cal F}}

\newcommand{\giro}[1]{\stackrel{\circ}{#1}}
\newcommand{\nc}{\giro{R}}
\newcommand{\mc}{R}
\newcommand{\nt}{\nc\hspace{-1mm}{}_t}
\newcommand{\mt}{\mc_t}
\newcommand{\me}{E}

\newcommand{\Iaa}{I}           %Q (21)
\newcommand{\BNd}{\giro{\cal F}_{N,\,t}}    % 1+B_{t} N^{\dag}_{t}
\newcommand{\NdB}{{\cal F}_{N,\,t}  }   %  1+N^{\dag}_{t} B_{t}
\newcommand{\BNdt}{\giro{\cal F}_{N,\,t+1}}   % 1+B_{t+1} N^{\dag}_{t+1}
\newcommand{\NdBt}{{\cal F}_{N,\,t+1}}  %   1+N^{\dag}_{t+1} B_{t+1}
\newcommand{\BNdtm}{\giro{\cal F}_{N,\,t-1}  } % 1+B_{t+1} N^{\dag}_{t+1}

\newcommand{\be}{\begin{equation}}
\newcommand{\ee}{\end{equation}}

\newcommand{\reff}[1]{(\ref{#1})}

\begin{document}

\title{Chiral symmetry breaking and \\ quark confinement \\ in the nilpotency expansion of QCD}

\author{
  {\small Sergio Caracciolo}                       \\[-1.7mm]
  {\small\it Dipartimento di Fisica and INFN}      \\[-1.7mm]
  {\small\it Universit\`a degli Studi di Milano}   \\[-1.7mm]
  {\small\it via Celoria 16, I-20133 Milano, ITALY}                \\[-1.7mm]
  {\small\tt Sergio.Caracciolo@mi.infn.it},        \\[-1.7mm]
  {\protect\makebox[5in]{\quad}}
   \\
  {\small Fabrizio Palumbo}  \\[-1.7mm]
  {\small\it INFN -- Laboratori Nazionali di Frascati}      \\[-1.7mm]
  {\small\it P.~O.~Box 13, I-00044 Frascati, ITALY}       \\[-1.7mm]
   {\small\tt fabrizio.palumbo@lnf.infn.it}       \\[-1.7mm]
  \\
 {\protect\makebox[5in]{\quad}}  % To force authors' names to be written
				  %   vertically, one above another.
				  % (\author seems to put them side-by-side
				  %   if there is room.)
}

\maketitle

\thispagestyle{empty}

\begin{abstract}

%\vspace{0.3cm}

We apply to lattice QCD a bosonization method previously developed in which dynamical bosons are generated by time-dependent Bogoliubov transformations. %,  respecting all symmetries. 
The transformed action can be studied by an expansion in the inverse of the nilpotency index, which is the number of fermionic states in the structure function of composite bosons. When this number diverges the model is solved by the saddle-point method which has a variational interpretation. We give a stationary covariant solution for a background matter field whose fluctuations describe mesons.  In the saddle-point approximation fermionic quasiparticles exist which have quark quantum numbers. They are confined in the sense that they propagate only in pointlike color singlets.  Conditions for chiral symmetry breaking are determined, to be studied numerically, and a derivation of a mesons-nucleons action is outlined.
%\noindent Keywords:
\end{abstract}

\clearpage

%\vfill\eject

\renewcommand{\thefootnote}{\arabic{footnote}}
\setcounter{footnote}{0}

%\clearpage

%\nopagebreak

\section{Introduction}

The fundamental fields appearing in QCD, that is quarks and gluons, are confined, and we are able to observe directly only mesons and baryons. It is conceptually interesting to see how these composite fields emerge from the microscopic dynamics and in several cases can be practically convenient to reformulate QCD in terms of them. The description of  the phase diagram of the theory, in particular,  should be more transparent in terms of these fields. We will refer to a reformulation of QCD in terms of hadronic fields as to QCD hadronization. 

This task goes beyond the perturbation theory in the gauge coupling constant, thus our starting point will be the lattice formulation of QCD~\cite{Wilson}.

After Wick rotation, in the Euclidean formulation, the path-integral of the pure-gauge sector could be studied using a number of tools, first of all Monte Carlo simulations, which have helped us to understand what occurs in the non-perturbative regime.
Within this framework, the role of quark fields soon appeared to be more difficult to consider. There is a fundamental difficulty with a lattice action for fermions which explicitly preserves locality, gauge and chiral symmetries: the duplication of the spectrum (see for example the textbook~\cite{MM}).
But also, as in the path-integral formulation fermionic fields are represented by Grassmann variables in a Berezin integral, an efficient numerical simulation seems to require the preliminary integration of the fermionic degrees of freedom. The recovered functional determinant is heavily demanding from the numerical point of view and is still a hard problem in the region of finite chemical potential because it is not positive definite. This prevents the definition of a probability measure, thus making the introduction of approximation schemes  difficult. 

A hadronization of QCD could overcome these difficulties, at least in the mesonic sector. We have, therefore, been pushed to apply to lattice QCD  a general method of bosonization we developed in recent years, both in Many Body and Field Theory, in the presence of fundamental fermions. It turns out that not only does such a method allow us to introduce mesons, but also, in some approximations, baryons.

Our starting point is  the operator formulation of the partition function in the Fock space representation for the fermionic fields, in which approximations can be introduced by following physical insight and can be mathematically justified.
In a first approach~\cite{Palu05, Cara} we restricted the evaluation of the partition function of a system to states of composite bosons, in a variational spirit. The resulting bosonic effective action in such restricted space was evaluated exactly. 

In such a framework a perturbative scheme was formulated  as follows. The composite bosons are characterized by an integer, the index of nilpotency, which is the number of fermionic states in their structure function. This is also the maximum number of composites which can exist in a given quantum state and therefore only when this number diverges can composite bosons  behave as canonical bosons~\cite{Cara}. We set a perturbative expansion in the inverse of this number which we call nilpotency expansion. Since the index of nilpotency counts the total number of independent fermionic modes in the composites, it is in general much greater than the number of internal degrees of freedom of the fermions. For instance the number of degrees of freedom of the electron is 2, but the total number of fermionic states in the Cooper pairs in superconductors is infinite in the thermodynamic limit, which is the reason why the BCS solution gives the exact energy per particle in this limit.

It was later recognized~\cite{Palu} that this variational approach provides the saddle-point approximation in the nilpotency expansion to a theory obtained from the original one by time-dependent Bogoliubov transformations. Bogoliubov transformations are a natural way to introduce composites as Cooper pairs~(see for instance~\cite{Misha} for a different approach along similar ideas, and~\cite{Marchetti} for the solution, directly in the continuum, of the problem of Dirac particles in an external  stationary magnetic field in 2+1 dimension by means of a series of Bogoliubov transformations). 
%Their application to relativistic quantum field theories is not a novelty~\cite{Arak}.  
But the specific difficulties of the renormalization procedure in the Hamiltonian formalism have limited their use in this domain. A more severe drawback in an application to gauge
theories is that symmetric terms of the original theory, in general, give rise to a sum of terms in the transformed actions  none of which conserves in general  the symmetries (even though, since Bogoliubov transformations are unitary,  all symmetries are conserved in the sum). This becomes a potential source of problems when some approximations are performed. 

In our approach  we try to avoid difficulties with renormalization in the Hamiltonian formalism  by   use of the transfer matrix formalism, which is at our disposal since we are using the lattice formulation.  
%The problem of term by term symmetry conservation was solved in the following way. 
By using this formalism we can oscillate between the operator formalism in Fock space and the functional formalism.

An independent  Bogoliubov transformation at each time slice is performed in the operator form of the partition function.
The time-dependent parameters of the transformation are required  to vary under symmetry transformations in such a way that the quasiparticles fields transform in the same way as the original fermion fields under the symmetries of the theory, in particular gauge invariance. These   parameters can then be associated with dynamical bosonic (composite) fields in the presence of fermionic fields (quasiparticles) with the quantum numbers of the bare fermions. 
A  compositeness condition  avoids double counting~\cite{Palu}. One thus gets an effective action of composite fields plus quasiparticles, exactly equivalent to the original one, in which ground 
and excited states can be treated on the same footing.  Of course, in practical applications, some approximation must be introduced.

As usual different solutions in a saddle-point approximation can be related to different phases of the theory and the nilpotency expansion will be our tool to study the dynamics of the composites.
There is a complete arbitrariness regarding the composites introduced by the Bogoliubov transformations because, since they are unitary, the transformed theory is  exactly equivalent to the original one irrespective of their choice. 
But only when the introduced composites reproduce the effective degrees of freedom in the given phase, the transformed theory, after suitable approximations, can have practical applications. 

In a successive investigation~\cite{Cara1} of our method, we looked into the nature of the saddle-point equations. In the saddle-point approximation  the Bogoliubov transformation on the transfer matrix has the same effect as the Foldy-Wouthuysen transformation on the Dirac Hamiltonian: it eliminates the direct mixing between fermions and antifermions.
In the absence of
gauge fields we found an explicit solution of the saddle-point equations~\cite{Cara}.
%for fermions regularized according to the Kogut-Susskind prescription~\cite{Cara}.

In the present work we consider the saddle-point equations in the presence of gauge fields, namely the application of our method to QCD, restricting   ourselves  to the case of zero chemical potential. A preliminary discussion of finite baryon density can be found in~\cite{Lattice2010}.  We use the Kogut-Susskind regularization for fermions in the flavour basis. The corresponding expressions in the spin-diagonal basis which is more commonly used in numerical simulations will be reported elsewhere.

We  find an exact covariant solution which requires that the vacuum should be dominated by stationary chromomagnetic fields. At the present stage of our research such a dominance appears to be driven by dynamical quark fields, but it should be remembered that dominance of chromomagnetic fields was already found and discussed by several authors for pure-gauge theories~\cite{Savvidy,Nielsen}. Because of the coupling with quarks,  the  QCD vacuum appears to be a dual 
superconductor (not a color superconductor). 
Dominance of chromomagnetic fields  in the non-Abelian case,  is indeed reminiscent of  the picture of color confinement based on electric-magnetic duality~\cite{Nambu,Pari,Mande,Thooft,Suganuma} and has been found also in the standard theory of color superconductivity~\cite{Kogu}.
This should be compared with the abelian case, where, by contrast, we do not expect that the configurations of gauge fields giving rise to a stationary nonvanishing magnetic field should dominate the vacuum.

Our vacuum contains a condensate of quark-antiquark pairs,
but these Cooper pairs do not have the quantum numbers of chiral fields, which are instead associated with the fluctuations of this condensate. For this reason we will refer to this condensate as to the background field. 

Does this background solution describe what we expect as essential features of QCD like confinement and chiral symmetry breaking?
Our answer can be only partially positive. 
 Concerning the first issue we obtained an important result:   in the background of a stationary chromomagnetic field  quasiparticles cannot propagate separately: only color singlet composites of quasiparticles can propagate and  therefore  have a physical particle interpretation. This means that color is confined as far as quasiparticles are concerned.  The quasiparticles color singlets can be mesons and baryons as well. Therefore not only can our method describe bosonization, it can also account for the more complicated process of formation of  composites of an odd number of quasiparticles. Concerning spontaneous chiral symmetry breaking instead we we have found conditions similar to the standard ones which must ultimately be solved numerically.

The expression of the background field is the starting point of our program of hadronization. This will require, at the present stage, the evaluation by numerical simulations (possible because of our lattice regularization) of some quantities appearing in the nilpotency expansion. An illustrative example of such quantities can be found  in Section~\ref{8}, where a derivation of a meson-nucleus action is outlined, and gauge-invariant coefficients depending on link variables appear explicitly. Also, the expression of the partition function at finite chemical potential that we found~\cite{Lattice2010}, which avoids the sign problem, requires an integration over spatial gauge link variables.

We wish to remark that our method is well suited to also study the effect of an intense background magnetic field on strong interactions, a problem considered of interest both at the level of the cosmological electroweak phase transition and for the heavy-ion collisions. And, indeed, numerical simulations have already been performed both in the quenched approximation, see for example~\cite{Poli}, and with dynamical fermions (see also~\cite{Delia}  for a detailed bibliography), in order to try to understand magnetic catalysis, i.e. the increase of chiral symmetry breaking induced by the magnetic background field.

The paper is organized in the following way. 
In Section~\ref{definitions} we establish our definitions and notations. To make the paper self-contained we report in Section~\ref{tdbt} the derivation of the effective action by  time-dependent  Bogoliubov transformations. 
In Section~\ref{sp} we report our solution of the saddle-point equations for gauge theories.
In Section~\ref{symmetries} we discuss the physical interpretation of our results and their relevance to symmetries, in particular chiral symmetries.
In Section~\ref{confinement} we prove confinement of quasiparticles in the saddle-point approximation.
In Section ~\ref{8} we outline a derivation of mesons-nucleons action from QCD.
Finally, in Section~\ref{end}, we perform a summary and give an outlook of our method.

\section{Definitions and notations\label{definitions}}

Consider a system of fermions interacting  with  external bosonic fields 
including gauge fields, regularized on a lattice. The
fermionic part of the partition
function at finite temperature $T$  can be written
\begin{equation}
\mathcal{Z}_F = {\Tr}^{F}  
\prod_{t=0}^{L_0/s-1} {\cal T}_{t,t+1} \,.
\label{part}
\end{equation}
 $L_0 =  T^{-1}$  is the number of links in the temporal direction, $\cal T$
is the fermion transfer matrix,   ${\Tr}^F$ is the trace  over the Fock space of fermions.
% and $M$ a matrix to be defined below. 
The parameter $s$ takes the value $1$ in the Wilson formulation for lattice fermions, but $s=2$ for the Kogut-Susskind fermions which live on blocks of  twice the size of the lattice spacing. The index $t$ labels  the blocks along the ``time" direction.

For Wilson fermions the expression of $\cal T$ was given by L\"uscher~\cite{Lusc},  in the gauge
$U_0 = \bone$, in which  one has to impose the Gauss constraint in the Hilbert space of the system (a Fock space of fermions in which the coefficients of the fermionic states are  polynomials of 
spatial link variables). One can also use 
a slightly modified form which avoids the Gauss constraint by reinstating the temporal links variables: 
\be
{\cal T}_{t,t+1} :=   e^{\frac{1}{2}\tr (M_t + M_t^{\dagger})} \,
\hat{T}^\dagger_t  \, {\hat V}_t \, e^{s\,\mu \, \hat{n}}\,  \hat{T}_{t+1} \,
\ee
where $\mu$ is the chemical potential and  $\hat{n}$ is the fermion number operator 
\be
\hat{n} :=  \hat{u}^{\dagger} \hat{u} - \hat{v}^{\dagger} \hat{v} \, ,
\ee
(the sum on all fermion  indices is understood) with $\hat{u}^\dagger$ and $\hat{v}^\dagger$ (respectively $\hat{u}$ and $\hat{v}$), creation (respectively annihilation)
operators of fermions and antifermions, obeying canonical anticommutation relations and 
\begin{eqnarray}
\hat{T}_t  &=&\exp [ -\hat{u}^{\dagger} M_t\, \hat{u} - \hat{v}^{\dagger}  
M^T_t \hat{v} ] 
\exp[\hat{v} N_t  \, \hat{u}] \\
{\hat V}_t &=& \exp  [ \hat{u}^{\dagger}\ln  U_{0,t}\, \hat{u} + \hat{v}^{\dagger}  \ln
U_{0,t}^* \, \hat{v} ] \,.
\end{eqnarray}
The matrices $M_{t}$ ($ { M}_t^T$ being the transposed of $M_t$) and $N_t$ are functions of the  spatial 
link variables at time $t$ and possibly of other  bosonic fields, such as the external magnetic fields considered in~\cite{Poli, Delia}.

Remarkably, Kogut-Susskind fermions, but in the so-called {\em flavour basis} (see for example~\cite{MM}), give rise to a transfer matrix of the same form~\cite{Palu-KS}.
Explicit expressions for  Wilson and Kogut-Susskind fermions in the flavor basis are  reported 
in  Appendix~\ref{transfer}.
The variables $ U_{0,t}$ are matrices in a unitary representation of the gauge group whose  elements are the link variables between Euclidean time $t$ and $t+1$
\begin{equation}
(  U_{0,t} )_{{ \bf x},{ \bf y} }
= \delta_{{ \bf x},{ \bf y} } \, U_{0}(t,{\bf x})
\end{equation}
where boldface letters, such as $\mathbf{x}$, denote spatial coordinates.

We introduced the following notation, which we will use  for any matrix $\Lambda$
\begin{equation}
\mbox{tr}_{\pm} \Lambda := \mbox{tr}\left( P_{\pm} \Lambda \right) \,.
\end{equation}
The  operators $P_{\pm}$  project on the  components of the fermion field which propagate forward or backward in time
\begin{eqnarray}
\hat{u} &=& P_{+} \hat{\psi}
\nonumber\\
\hat{v}^{\dagger} &= &P_{-} \hat{\psi}
\end{eqnarray}
and their expressions are given in Appendix~\ref{transfer}. The symbol
 ``$\mbox{tr}$" denotes  the trace over fermion-antifermion intrinsic quantum numbers and spatial coordinates (but not over time).
Finally we will denote by $T_0^{(\pm)}$  the forward and backward
translation operators of one block, that is $s$ lattice spacing, in the ``time"  direction 
\begin{equation}
[T_0^{(\pm)}]_{t_1,t_2 }=\delta_{t_2,t_1 \pm 1} \,.
\end{equation}
The expression of the full partition function is
\be
{\mathcal Z} = \int [dU] \exp (-S_G) \, {\mathcal Z}_F
\ee
where $S_G$ is the pure gluon action.

\section{Time-dependent Bogoliubov transformations\label{tdbt}}

The material of this Section is taken from Ref.~\cite{Cara1} and is reported to make the paper self-contained.

Usually  the trace appearing in the definition of the transfer matrix is evaluated using at each time slice coherent states of fermions
\begin{equation}\label{coherent}
| \alpha,\beta\rangle = \exp (- \alpha \, {\hat u}^{\dagger} - \beta \, {\hat v}^{\dagger} ) | 0 \rangle,
\end{equation}
where the $\alpha ,\beta$ are Grassmann fields.
 We will use instead states
obtained by applying, at each time slice, an independent  Bogoliubov transformation,  
\be
\hat{\psi}_{\mathcal F} = \left[ P_+ \, R^{\frac{1}{2}} ( 1 - {\mathcal F}^{\dagger} ) + P_- \giro{R}{\hspace{-1mm}}^{\frac{1}{2} } ( 1 + {\mathcal F} ) \right] \, \hat{\psi} \label{Bogoliubov}
\ee
where
\begin{equation}
 R= (1 + {\mathcal F}^{\dagger} {\mathcal F})^{-1} \qquad  \giro{R} \, = (1 + {\mathcal F} {\mathcal F}^{\dagger})^{-1} 
\label{involution}
\end{equation}
and $ {\mathcal F}$ is an arbitrary matrix such that
\be
P_\pm  {\mathcal F} =   {\mathcal F} P_\mp \, .
\ee
The circle over the $R$ denotes the involution
defined by the above equations.  The new operators (we omit the subscript ${\mathcal F}$ to lighten the notation)
\begin{align}
\hat{\alpha} = & \, P_+ \hat{\psi}_{\mathcal F} = R^{\frac{1}{2}}\left( {\hat u} -   \,{\mathcal F}^{\dagger}  \, {\hat v}^{\dagger}\right) \label{Ba} \\
{\hat \beta}^{\dagger} = & \, P_- \hat{\psi}_{\mathcal F} =  \, \giro{R}{\hspace{-1mm}}^{\frac{1}{2} } \left( {\hat v}^{\dagger} + {\mathcal F}\,{\hat u}\right)    \label{Bb}
\end{align}
satisfy canonical commutation relations for any choice of the matrix $ {\mathcal F}$. We will let $ {\mathcal F}$ depend on all the fields coupled to the fermions in such a way as to respect as many symmetries as we can. The vacuum of the new operators is
 \be
  |\mathcal{F} \rangle= \exp (\hat{\mathcal{F}}^{\dagger}) \,|0 \rangle \label{defF}
  \ee
  where
\begin{equation}
{\hat \fc }^{\dagger} = \uh^\dag \fc^\dag  \vh^\dag, \label{compbos}
\end{equation}
is a the creation operator of a composite boson. As already mentioned the new vacuum appears as a coherent state of  fermion-antifermion pairs.
We also remark that the new vacuum defined in~\reff{defF} is gauge-invariant and therefore satisfies the Gauss constraint.
The transformed states can be written
\begin{align}\label{bogcoherent}
 \hat{\mathscr U}(\mathcal{F})\, | \alpha,\beta\rangle = & \, (\mbox{det}_+\, R^{\frac{1}{2}})\, | \alpha,\beta; \fc\rangle  \\
 = & \, (\mbox{det}_+\, R^{\frac{1}{2}})\,\exp (- \alpha \, {\hat \alpha}^{\dagger} - \beta \,{\hat \beta}^{\dagger} ) |\mathcal{F} \rangle \\
= & \, (\mbox{det}_+\, R^{\frac{1}{2}})\, \exp \left( {\hat{\cal F }}^\dag - a\,   \alphah^\dag - b\,  \betah^\dag - \beta {\cal F } \alpha \right) \big|0 \rangle 
\end{align}
where $\mbox{det}_\pm$ is the determinant in the subspace where $P_\pm$ projects, with 
\begin{align}
\mbox{det}_+\, R  =  &\, \mbox{det}_-\, R \\
\mbox{det}_\pm\, R = &\, \mbox{det}_\pm \giro{R}
\end{align}
and  $a:=\mc^{-\meta}\alpha$ and $b:=\beta \giro{R}{\hspace{-1mm}}^{-\frac{1}{2}}$. The explicit definition of the operator $ \hat{\mathscr U}$ can be found in~\cite[Appendix B]{Cara1}, and here we correct a misprint in~\cite[(2.16-2.18)]{Cara1} where the normalization factor $\mbox{det}_+\, R^{\frac{1}{2}}$ had been forgotten.

After evaluation of the trace,  the partition function becomes
\begin{eqnarray}\label{Z}
{\mathcal Z}_F &=\exp\{-S_{me}( {\mathcal F}) \}
\int{D[\alpha^*,\alpha,\beta^*,\beta]\,  \exp\{ -S_{qp}(\alpha,\beta;\fc) }\}
\end{eqnarray}
where the Grassmann variables $\alpha^*, \alpha, \beta^*, \beta$ satisfy antiperiodic boundary conditions in time.
In the above equation $S_{me}$, the  term independent of the Grassmann variables,
 will  be interpreted as a meson action
\be\label{bosonaction}
   S_{me}(\fc) := - \, \sum_{t=0}^{L_0/s-1}\tr_{+} \ln\left ( R_t \,U_{0,t}\, \me_{t+1,t}  \right) = - \, \sum_{t=0}^{L_0/s-1}\tr_{+} \ln\left ( R_t \, \me_{t+1,t}  \right) 
\ee
where
\be
 \me_{t+1,t} :=   \,\left(\NdBt\right)^\dag \eMut \,e^{M_t^\dag}\,\NdB
 +\, \fc_{t+1}^\dag\,\emenMvt \, e^{-M_t^\dag}\, \fc_{t}\,,  
\ee
with
\be
 \NdB :=1+N^{\dag}_{t} \fc_{t}\,
\ee
and we used the fact that $U_{0,t}$ are unitary.

The other term is the action of quasiparticles 
\begin{multline}\label{SF}
S_{qp}(\alpha,\beta;\fc)=  - s \sum_{t=0}^{L_0/s-1}\Big[ \beta_{t+1} {I }_{t+1}^{(2,1)} \alpha_{t+1} + \alpha^*_{t}{I }_t^{(1,2)} \beta^*_{t}\\
	 + \alpha^*_t  (\nabla_t -{\cal H}_t) \alpha_{t+1}-\beta_{t+1}  (\giro{\nabla}_t-\giro{{\cal H}_t)}\beta^*_{t}\Big]  
\end{multline}
written in terms of lattice covariant derivatives
\begin{align}
\nabla_t&:=s^{-1}\left(e^{s\mu}U_{0,t}-T^{(-)}_0\right) \\
\giro{\nabla}_t&:=s^{-1}\left( e^{-s\mu}U_{0,t}^\dag -T^{(+)}_0\right)
\end{align}
and the lattice Hamiltonians, respectively, for fermions and antifermions,
\begin{align}
{\cal H}_t \,:= &\, s^{-1}e^{s\mu}\left(  U_{0,t} - \mt^{-\frac{1}{2}}\me_{t+1,t}^{-1}\mc_{t+1}^{-\frac{1}{2}} \right) \\
\giro{{\cal H}}_t\, :=& \,  s^{-1}e^{-s\mu}\left( U_{0,t}^\dag - \nc\hspace{-1mm}{}^{-\frac{1}{2}}_{t+1} \giro{\me}\hspace{-1mm}{}_{t+1,t}^{-1}\nt^{-\frac{1}{2}}
\right)  \, .
\end{align} 
There are, in addition,  unwanted terms which mix  quasiparticles with quasiantiparticles whose coefficients are
\begin{align} 
  \Iaa_t^{(2,1)} & := 
 \, s^{-1}\,  \nt^{\frac{1}{2}}\,  \left[ \nt-\giro{\me}\hspace{-1mm}{}_{t,t-1}^{-1}\BNdtm ~e^{M_{t-1}^{\dagger}}\eMvtmenone  \right] \fc_{t}^{\dag -1}  \mc_{t}^{\frac{1}{2}} \\
 \Iaa_t^{(1,2)} &:= 
  \, s^{-1}  \mc_{t}^{\frac{1}{2}}\,  \fc_{t}^{-1}\left[ \nt-  e^{M_{t}^\dag}~\eMvt~\left(\BNdt\right)^\dag \giro{\me}\hspace{-1mm}{}_{t+1,t}^{-1}\right]\, \nt^{\frac{1}{2}} \, .
\end{align}
The definitions of the other new symbols are
\begin{align}
 \giro{\me}_{t+1,t}\,  := &  \,  \BNd \,e^{M_t^\dag}\,\eMvt\,\left(\BNdt\right)^\dag
	  +\,  \fc_{t}\,e^{-M_t^\dag}\,\emenMut\fc_{t+1}^\dag \\
  \BNd \,  := &  \, 1+\fc_{t} N^{\dag}_{t} \, .
\end{align}

\section{Saddle-point equations and factorization of the transfer matrix\label{sp}}

We assume that the contribution of quasiparticles to the vacuum energy be negligible. Therefore in order to determine the contribution of the fermions to the vacuum energy we must minimize the mesonic action with respect to ${\mathcal F},{\mathcal F}^{\dagger} $. This gives  the  saddle-point equations, valid for $ 0 \le t \le \frac{L_0}{s}-1$
\begin{align}
\fc_{t+1}\,= \,& N_{t+1} +\emenMvt  e^{-M_t^\dag} \fc_t \big(\fc_{N,\,t}\big)^{-1} e^{-M_t^\dag} \emenMut  \\
 \fc_{t}^\dag\,= \,& N_{t}^\dag + e^{-M_t^\dag}\emenMut  \big(\fc_{N,\,t+1}^\dag\big)^{-1} \fc_{t+1}^\dag \emenMvt e^{-M_t^\dag}\,. \label{saddle}
\end{align}

The main difficulty of the saddle-point equations stems from their dependence on time. This difficulty is reduced if we look for stationary solutions, as appropriate to the vacuum.  If ${\mathcal F}$ is stationary,  the elementary bosonic fields coupled to the fermions which enter its expression should also be stationary~\cite{Cara}.  In gauge theories ${\mathcal F}$ certainly depends on spatial link variables. Stationarity for  gauge fields can be formulated in a gauge covariant way by requiring  that these fields evolve according to gauge transformations, so  we must require that
 \be
U_{k} (t, {\bf x}) = W^{\dagger}_{t, {\bf x}} U_{ k }(0,{\bf x}) W_{t, {\bf x}+{\hat {\bf k}}} \, . \label{ansatz}
\ee
As a consequence the chromomagnetic contribution to the pure gauge-field action, namely the trace of  spatial plaquettes, does not depend on time.

Accordingly, the matrices $N_t, M_t$ are related to those at time $t=0$,  that is, if  $N_0 = N$ and  $M_0 =  M$, by
 \be
 N_t  = W^{\dagger}_t  \,  N \, W_{t}\, , \qquad M_t  = W^{\dagger}_t  \, M \, W_{t} \, .
\ee

We still wish to set the contribution of the chromoelectric field  to the gauge action, namely the trace of spatio-temporal plaquettes, to be independent on time.
We have been able to arrive at a stationary solution for $\mathcal F$   only with the particular choice
\be
W_{t+1, {\bf x}} = U_{0}(0,{\bf x})\, U_{0}(1,{\bf x}) \dots U_{0}(t,{\bf x})
\ee
which lets the contribution from the chromoelectric field vanish at all times. 
Indeed, if we do not consider colored composites, 
the saddle-point equations for  
 \be
 {\mathcal F}_t  = W^{\dagger}_t  \,  {\mathcal F}_0  \, W_{t} = W^{\dagger}_t  \,  {\mathcal F}  \, W_{t}
\ee
then become 
\be
{\mathcal F} =  N + e^{-M}  e^{-M^\dagger}{\mathcal F} \, {\mathcal F}_{N}^{-1}
 e^{-M^\dagger}  e^{-M}
\ee
and the Hermitian conjugate relation.   

Assuming $ M=M^{\dagger}$, a condition satisfied in both the Wilson and Kogut-Susskind regularizations,  under the condition $[N,M]=0$ (which is not satisfied by the Wilson regularization in presence of a nontrivial gauge configuration),  their solution~\cite{Cara, Cara1} is
\begin{equation}
\overline{{\mathcal F}}= N(2N^{\dagger}N)^{-1}  \left[-Y
  + \sqrt{ Y^2 + 4  N^{\dagger}N} \right] \,. \label{Fsaddle}
\end{equation}
where
\begin{equation}
Y=1- N^{\dagger}N - e^{-4M}\,.
\end{equation}

The time evolution of  the quasiparticle Hamiltonians is slightly different 
\be
{\mathcal H}_t=  W^{\dagger}_t  \,  {\mathcal H} \, W_{t+1}\, ,
\qquad 
\giro{\mathcal H}_t=  W^{\dagger}_{t+1}  \,  \giro{\mathcal H} \, W_{t}\, .
\ee

At the saddle point (we will overline all quantities evaluated at the saddle point)
\be
 e^{-s \mu} \, {\overline {\mathcal H}} = e^{s \mu} \, \giro{{\overline {\mathcal H}}}\, = \frac{1 }{ s} \,  
\left[ 1- \, {\overline {\mathcal F}}_N^{-1}  e^{- 2 M} \right] \, ,
\ee
so that ${\overline {\mathcal H}}$ and $\giro{{\overline {\mathcal H}}}$ are  Hermitian functions of $M$ and $N^{\dagger} N$ and the vacuum energy is
\be
\overline{S}_{me} = S_{me} ({\overline {\mathcal F}}) =  - \frac{L_0}{s}  \, \mbox{tr}_+ \ln \overline{Q}
\ee
where we introduced 
\be
\overline{Q} =  \left( 1 - s\, e^{-s \mu} \, {\overline {\mathcal H}} \right)^{-1} \label{Q}
\ee
for future convenience.

For Wilson fermions we have not been able to find an exact solution, but in the following we will assume its existence.

Since the matrices $N,M$  are Hermitian and by assumption commute with each other, they can be diagonalized simultaneously and ${\overline {\mathcal F}} $ is diagonal in such a basis. Labeling each eigensubspace by the index by $i$ and denoting by ${\overline {\mathcal F}}_i $ the corresponding eigenvalue,  for each state we can choose either to perform the Bogoliubov transformation by using the solution~${\mathcal F}_i ={\overline {\mathcal F}}_i $ or to leave the subspace unchanged by choosing ${\mathcal F}_i =0$. At zero temperature and chemical potential the first choice minimizes the vacuum energy, but
increasing the chemical potential because of Pauli blocking for an increasing number of states we must make the second choice. This is the mechanism for chiral
symmetry restoration found in a four-fermion interaction model~\cite{Palu, Cara1}, and confirmed for gauge theories in a forthcoming paper~\cite{Cara2}.

The effective mesonic action and therefore the saddle-point approximation can be obtained  also by  a variational calculation~\cite{Cara1} in which we  assume as a test fermionic state  the quasiparticle vacuum $|{\mathcal F}\rangle$.  We then  fix  the gauge according to $U_0(t) =1$. In the presence of such a gauge fixing we must impose the Gauss constraint in the Hilbert space. But since the state
$|{\mathcal F}\rangle$ satisfies the Gauss constraint  by construction,  we do not need to think about it any longer. Under such conditions the remaining gauge fields are independent of time and therefore automatically satisfy periodic boundary conditions in the time direction.

\subsection{Background field and vacuum properties}

It might at first sight be puzzling that the form of the saddle-point solution does not depend on whether the theory is or is not Abelian. This point requires some discussion.
 
 We first observe that  the saddle-point equations are identical to the conditions
\be
 		\Iaa{}_t^{(2,1)}\,=\,\Iaa{}_t^{(1,2)}=0\,.
\ee
Therefore in the saddle-point approximation  the fermion-antifermion mixing disappears in the quasiparticles action:
the Bogoliubov transformations~\reff{Bogoliubov} in the saddle-point approximation  factorize the transfer matrix into a term for quasiparticles and a term for antiquasiparticles~\cite{Cara1}. Hence their effect is analogous to the Foldy-Wouthuysen transformation which separates positive from negative energy states in the Dirac Hamiltonian. With respect to this factorization there is no difference between Abelian and non-Abelian theories. The role of the condensate ${\hat {\mathcal F}}$ is only to provide the background in which quasiparticles and antiquasiparticles propagate independently, and, as we will see in the next Section, it does not have a particle interpretation. For this reason we call ${\mathcal F}$ at the saddle point a background field.

In order to proceed with our analysis we must  distinguish two cases. In the first the vacuum is dominated by  chromomagnetic fields with nonvanishing energy, while in the second  the dominating fields are pure gauge fields. Here we expect a drastic difference beteen Abelian and non-Abelian theories, because  we think that the first/second case is realized in the continuum limit of non-Aabelian/Abelian gauge theories. Then in non-Abelian gauge theories because of the nontrivial gauge-invariant vacuum,
 temporal link variables disappear from both the gauge-field  and mesonic actions, but not from the quasiparticle action:  the QCD vacuum in the saddle-point approximation appears as a dual superconductor (not color superconductor)  which expels chromoelectric fields altogether  (in this   dual Meissner effect the penetration length vanishes). Fluctuations of chromoelectric fields  are subdominant, and as consequence we will see in Section~\ref{confinement} that quasiparticles are confined. In the Abelian case on the contrary,  fluctuations of gauge fields dominate and there is no confinement of quasiparticles.

\section{Symmetries and compensating fields~\label{symmetries}}

In this Section we shall consider only transformations $s$ associated with symmetry groups which act in a unitary linear representation on the fermionic field
\be
\hat{\psi} \to \hat{\psi}' = s \, \hat{\psi}
\ee
and leave  the action invariant.

Since Bogoliubov transformations are unitary they preserve such symmetries. 
But individual terms of the original action which are invariant, are transformed, in general,  into terms which do not enjoy  this property any longer, and symmetry conservation of the transformed total action is realized through  compensations among such non invariant terms. Then there is the danger that approximations can disrupt such compensations resulting in effective symmetry breaking.

This drawback can be  avoided in many cases, requiring that the quasiparticle fields should transform in the same way as the original fermionic fields, because then  invariant terms would obviously be transformed into invariant terms. This can be achieved by making the Bogoliubov transformations at each time slice dependent on time, and introducing, when necessary, compensating fields.  

We will restrict ourselves to symmetries 
which do not mix the components which propagate forward and backward in time, that is,
\be
[s, P_\pm] = 0 \,, \label{mix}
\ee
so that
\be
{\hat u}' = s \, {\hat u} \,, \qquad ({\hat v}')^{\dagger}  = s \,{\hat v}^{\dagger}  \,. \label{u'}
\ee
Then the quasiparticle operators change according to
\begin{align}
\hat{\alpha}'  =& \, s\, \left( s^{\dagger} R^{\frac{1}{2}}  s\right) \left( {\hat u} - \left(s^{\dagger} {\mathcal F} s\right)^{\dagger}  {\hat v}^{\dagger} \right)\\
( \hat{\beta}' )^{\dagger} =& \, s\,\left( s^{\dagger} \giro{R}{\hspace{-1mm}}^{\frac{1}{2} } s\right) \left( {\hat v}^{\dagger} + \left(s^{\dagger} {\mathcal F} s\right) {\hat u} \right)
\end{align}
namely they are still defined by the action of $s$ on  a Bogoliubov transformed field where instead of $\mathcal F$  the modified matrix
$ s^{\dagger} \,{\mathcal F}  s$ is used.
A simple way to preserve the symmetries as in the starting action is recovered if  we require that  the ${\mathcal F}$-matrix changes under the symmetry transformation according to
\be
{\mathcal F}' = s\, {\mathcal F} s^{\dagger} \,. \label{F'}
\ee
In order to enforce the above condition let us expand the matrices ${\mathcal F}_t $ at a given time-slice in the basis of time-independent matrices $\Phi(K)$ labeled by the indices $K$ (which also include  space):
\begin{equation}
{\mathcal F}_t  = \sum_{ K } \varphi^*_t(K)  \Phi (K) =  (\varphi_t  , \Phi )\, .
\end{equation}
Then \reff{F'} becomes
\be
 (\varphi'_t  , \Phi ) =  (\varphi_t  , s\,\Phi\,  s^\dagger )\, .
\ee
The transformation of the basis matrices can be written as
\be
s\, \Phi(K) \,s^{\dagger} =   \sum_{K'} S_{KK'}  \Phi(K')  = (S\,\cdot \Phi)(K)
\ee
so that
\be
 (\varphi'_t  , \Phi ) =  (\varphi_t  , S\,\cdot \Phi) = (S^\dagger \cdot \varphi_t  , \Phi)
\ee
which is to say that
it is necessary to require that  the  expansion coefficients transform according to 
\be
\varphi'_t(K)  = \sum_{K'} S_{KK'}^{\dagger} \varphi_t(K')  \,.
\ee
The above construction also provides  a physical interpretation of our formalism. Indeed we observe that, since we could perform a unitary transformation with an arbitrary ${\mathcal F}$-matrix and then an arbitrary expansion coefficients $\varphi(K)$, we can integrate over  them  with an arbitrary probability measure getting
\be
{\mathcal Z}_F =
\int{d\mu(\varphi) \, D[\alpha^*,\alpha,\beta^*,\beta]\, 
\exp\{-S_{me}({\mathcal F}) -S_{qp}(\alpha,\beta;\fc) }\}  \,.
\ee
Looking at the form~\reff{bosonaction} of $S_{me}$ we immediately realize that time derivative  terms are generated for the compensating fields. As a consequence, unless the basis matrices $\Phi(K)$ are invariant,  the expansion coefficients $\varphi(K)$ must  become  dynamical bosonic fields. The basis matrices then acquire the meaning of structure functions of mesonic composites with quantum numbers $K$. $K$ includes color for colored mesons, which should exist only in the deconfined phases. The choice of the basis matrices $\Phi(K)$ (whose form must be determined by a variational calculation) selects which mesons one will include in the calculation in a variational spirit.

We can look for an approximation to this expression of the partition function by determining the minimum of the action with respect to ${\mathcal F}^{\dagger}, {\mathcal F}$. The phases of the theory are determined by the solutions ${\overline {\mathcal F}}^{\dagger}, {\overline {\mathcal F}}$ of the saddle-point  equations.  
By construction if ${\overline {\mathcal F}}^{\dagger}, {\overline {\mathcal F}} $ are matrices which minimize the action, then the rotated matrices ${\overline {\mathcal F'}}^{\dagger}, {\overline {\mathcal F'}}$ must also be minima of the action. Therefore either they coincide with the unrotated matrices or the solution of the minima are degenerate. And this accounts for the breaking of the symmetry.

A perturbative expansion is realized by setting  
\be
{\mathcal F} = {\overline {\mathcal F}}^{\dagger} + \delta {\mathcal F}^{\dagger} \,,  \qquad
{\mathcal F} = {\overline {\mathcal F}} + \delta {\mathcal F}\,.
\ee
We assume the index of nilpotency of the structure functions  appearing in the fluctuations $ \delta {\mathcal F}^{\dagger}, \delta {\mathcal F}$ as an asymptotic 
parameter and perform an expansion in the inverse of the nilpotency number that we call the nilpotency expansion. 
The fluctuations $\delta {\mathcal F}^{\dagger}, \delta{\mathcal F} $ describe in the nilpotency expansion 
 interacting mesons of the form ${\hat u}^{\dagger} {\hat v}^{\dagger}, {\hat v} \,{\hat u}$, which we will call, for easy reference,  of ${\mathcal F}$-type, but not of the form ${\hat u}^{\dagger}  {\hat u}, {\hat v}^{\dagger}  {\hat v}$, which are not of ${\mathcal F}$-type. A discussion about some mesons which are not of ${\mathcal F}$-type is presented later in Section~\ref{8}. An example of such an expansion can be found in Ref.~\cite{Cara}.

We emphasize that from a mathematical point of view the new expression of the partition function is exactly equivalent to the original one. 
Note also that there is no double counting because the property of quasiparticles
to annihilate the vacuum
\begin{equation}
{\hat \alpha}_i |{\overline  {\mathcal F}} \rangle =  {\hat \beta}_i |{\overline {\mathcal F} } \rangle = 0
\end{equation}
can be interpreted as a
 compositeness condition: mesonic states are orthogonal to quasiquark-quasiantiquark states. This constraint has the physical meaning of the condition
$Z=0$ for bound states in the  Lehmann spectral representation of composite operators~\cite{Houa,Salam} (see also~\cite[Vol I, p. 461]{Weinberg}),
namely the  condition required to introduce a bound state on the same footing as the constituents
in a Lagrangian.

\subsection{Some examples}

We now give  a few illustrative examples. The first one concerns fermion number conservation in the nonrelativistic theory of many-body systems~\cite{Palu05}. We include it because it shows clearly the need and the physics of compensating fields and because it was the first application of our method. Moreover, the Bogoliubov transformation is, in this case, similar to that necessary in the relativistic theory of diquarks~\cite{Cara2}; namely,  it mixes  the annihilation and creation operators of one and the same fermion (quark, electron, nucleon, \dots),
\be
{\hat \alpha}  =  R^{\frac{1}{2}}\left( {\hat u} -  
\,{\mathcal F}^{\dagger}  \, {\hat u}^{\dagger}\right) 
% \, , \qquad
%{\hat \alpha}^{\dagger}  =  
%	\left( {\hat u}^{\dagger} - {\hat u} \, {\mathcal F}
%	\right)  R^{\frac{1}{2} }  
%	\nonumber\\
\ee
while in the relativistic case, see \reff{Ba} and \reff{Bb}, the transformation mixes the annihilation operator of a fermion with  the creation operator of the corresponding antifermion.
As a consequence the operator $\hat{\mathcal F}$ 
\be
\hat{\mathcal F} = \hat{u}\, {\mathcal F} \hat{u}
\ee
carries fermion number 2  instead of zero. 
Under the relevant symmetry  associated with fermion number conservation
\be
%{\hat u}^{\dagger} = e^{ - i \theta} \, ({\hat u}')^{\dagger}  \, , \qquad 
{\hat u}' = s\, {\hat u} = e^{ i \theta} \, {\hat u}
\ee
the ${\mathcal F}$-matrix transforms according to
%\be
%{\mathcal F}' =  s^*  \, {\mathcal F} s ^{\dagger} \,. 	
%\ee
%so that
\be
{\mathcal F}' = s^*\, {\mathcal F} s^{\dagger}  =    e^{-2i \theta} {\mathcal F}\,  .
\ee
Since in this case the structure functions $\Phi(K)$ can be taken invariant, we must require that
\be
\varphi'(K) = e^{  2i \theta} \varphi(K)\, .
\ee
Namely, we need compensating fields  $\varphi(K)$  with fermion number $2$. These fields describe the low energy excitations.

We come back now to the relativistic cases of composites of fermion number zero. The first  one concerns the results obtained by the application of our method to a four-fermion model~\cite{Cara}, which at zero mass enjoys a discrete chiral symmetry generated by the parity transformation
\be
s = - (\gamma_5 \otimes t_5) 
\ee       
which commutes with the projectors $P_\pm$.

The interaction can be bi-linearized by introducing an auxiliary bosonic field $\sigma$ coupled to the fermions according to Eq.~(\ref{etacoupling}).
The relevant matrix in the basis  is
\be
\Phi =  \gamma_0 \otimes \bone 
\ee
so that
\be
s\, \Phi\, s^\dagger = - \Phi
\ee
and the compensating field must change sign under parity. This model is also interesting here because it shows that the compensating field is exactly the field $\sigma$ coupled to the fermions.

The second example deals with  the residual chiral symmetry with Kogut-Susskind fermions in the flavor basis.  For zero fermion mass the QCD action is invariant under the continuous chiral transformations 
\be
s =  \exp\left(- \frac{i}{2} \, \gamma_5 \otimes t_5 \, \theta  \right) 
\ee
parametrized by the angle $\theta$.

The ${\hat \sigma}$-field is
\be
{\hat \sigma}= {\hat \psi}^{\dagger} ( \gamma_0 \otimes \bone) {\hat \psi} = {\hat u}^{\dagger} ( \gamma_0 \otimes \bone) \, {\hat v}^{\dagger} + {\hat v} \, (\gamma_0 \, \otimes \bone) \, {\hat u} 
\label{sigmaK-S}
\ee
and the Goldstone pion which corresponds to the axial symmetry at $m=0$ is
\be
{\hat \pi}= i\,{\hat \psi}^{\dagger} (\gamma_0 \gamma_5 \otimes t_5) \,{\hat \psi}  = i\,   {\hat u}^{\dagger}   (\gamma_0 \gamma_5 \otimes t_5) \, {\hat v}^{\dagger} +
i\, {\hat v} \ (\gamma_0 \gamma_5 \,\otimes t_5 )\,  {\hat u}  \,. \label{piK-S}
\ee
We can  write
\be
{\mathcal F}_t = \sigma_t^*  \,  \Phi_{\sigma}+ \pi_t^*  \, \Phi_{\pi} 
\ee
where  the basis matrices are
\be
\Phi_{\sigma} =  \gamma_0 \otimes \bone \, , \qquad \Phi_{\pi} = i \,\gamma_0 \gamma_5 \,\otimes t_5  \,.
\ee
Under infinitesimal chiral transformations these basis matrices transform according to
\be
s \,\Phi_{\sigma}\, s^\dagger \approx  \Phi_{\sigma} + \theta \, \Phi_{\pi} \, ,  \qquad  s \,\Phi_{\pi}\, s^\dagger \approx  \Phi_{\pi}- \theta \, \Phi_{\sigma}\,.
\ee
Therefore the  compensating fields $ \sigma_t, \pi_t$ must transform in the inverse way. 

There are 15 more pions in four dimensions which can be constructed with the taste matrices $t$~\cite{Vers,Golt}. Only those of 
${\mathcal F}$-type can be described  by the fluctuations $\delta {\mathcal F}^{\dagger}, \delta{\mathcal F} $. For our illustrative purposes it is sufficient to
consider the Goldstone pion. 

The  most important  application concerns gauge invariance in  QCD. Let us consider the case of a gauge transformation.
In addition to (\ref{u'}), where $s$ is replaced by the gauge transformation $g(t,{\bf x})$, we have the transformation for the spatial link variables 
\be
U_k'(t, {\bf x})= g(t,{\bf x}) \, U_k(t, {\bf x})\, g^{\dagger}(t,{\bf x+ \hat{k}})\, .
\ee
If we concentrate on colorless $\hat{\mathcal F}$,  the matrix ${\mathcal F}_t$ will depend on color only through the configuration of spatial links $U_{k,t}$ 
\be
{\mathcal F}_t = {\mathcal F}(U_{k,t})
\ee
where the matrices $U_{k,t}$  are such that
\be
\left( U_{k,t} \right)_{\bf{x}, \bf{y}} =  \delta_{\bf{y}, \, {\bf x+ \hat{k}} }\, U_k(t, {\bf x})\, .
\ee
In this case, by also introducing  the matrices $g$ and using the matrix multiplication,  (\ref{F'}) becomes
\be
{\mathcal F}(U')= {\mathcal F}(g \, U \, g^{\dagger})= g \, {\mathcal F}( U) \,g^{\dagger} 
\ee
and it is automatically satisfied.
Therefore, as far as  gauge invariance is concerned, no compensating fields are needed.

In all the cases mentioned above  the effective action respects the original symmetry term by term and the quasiparticle vacuum is invariant provided we perform the symmetry transformations on the fermions and on the
compensating fields as well. Such a vacuum can be regarded as a condensate of the composites $\hat{{\overline {\mathcal F}}}$. We remark,
however, that  for Kogut-Susskind fermions these composites have a structure different from that of the chiral mesons,
\be
{\overline {\mathcal F}}\neq {\overline \sigma}^*     \Phi_{\sigma} +  {\overline \pi}^* \, \Phi_{\pi} 
\ee
and therefore
the quasiparticle vacuum cannot be interpreted as a condensate of these physical particles. 
%We will see that with Wilson fermions the vacuum is not a condensate of $\sigma$-mesons. 
In this case we will refer to the field  $\hat{\overline{{\mathcal F}}}$ as to a background field. But we should keep in mind that its fluctuations describe ${\mathcal F}$-type mesons which include the chiral ones.

\subsection{Spontaneous chiral symmetry breaking}

We will discuss spontaneous chiral symmetry breaking in the saddle-point approximation. 
We will not derive any new results. Our purpose is only to formulate this problem in our formalism.

To evaluate the order parameter, at fixed gauge configuration, we shall use the relation 
\be
\langle \bar{\psi} \psi \rangle_F = \frac{\partial}{\partial m} \log {\cal Z}_F 
\ee
and  our saddle-point approximation for the partition function, that is,
\be
\langle \bar{\psi} \psi \rangle_F = - \frac{\partial}{\partial m}  \overline{S}_{me} \, ,%= \frac{\partial}{\partial m} \tr_- \log \overline{E}
\ee
which, we remind the reader, is justified only for the special gauge configurations we discussed earlier. By direct calculation we get
%We then have in the saddle-point approximation
%\begin{eqnarray}
\be
\langle \bar{\psi} \psi \rangle_F =
%\langle  \overline{{\mathcal F}} | \hat{\sigma} |   \overline{{\mathcal F}}  \rangle  &= & %\approx \langle {\overline {\mathcal F}}| {\hat \sigma}_{\mbox{antinormal}} |  {\overline {\mathcal F}} \rangle 
%\tr_+  {\overline R} \left[ {\overline {\mathcal F}}^{\dagger}(\gamma_0 \otimes \bone) +(\gamma_0 \otimes \bone) {\overline {\mathcal F}} \right]
% + 
%{\langle  \overline{{\mathcal F}} |   \overline{{\mathcal F}}  \rangle}  
\, \tr_-   {\giro {\overline R}} \left[( \gamma_0 \otimes \bone) \, {\overline {\mathcal F}}^{\dagger}
+ {\overline {\mathcal F}}\, (\gamma_0 \otimes \bone) \right]
%\nonumber\\
%\langle  \overline{{\mathcal F}} | \hat{\pi} |  \overline{{\mathcal F}}  \rangle &=& 0\,.
%\end{eqnarray}
\ee
and by substitution of ${\overline {\mathcal F}}$ 
\be
%\frac{\langle  \overline{{\mathcal F}}  | \hat{\sigma} |  \overline{{\mathcal F}}  \rangle} {\langle  \overline{{\mathcal F}} |   \overline{{\mathcal F}}  \rangle}  
\langle \bar{\psi} \psi \rangle_F
=  - \,2\, m \, \tr_- \left[  \frac{1}{H  \sqrt{1 +H^2} }\right]  =  - \, m \, \tr \left[  \frac{1}{H  \sqrt{1 +H^2} }\right]\,
\ee
%{\bf mancano dei fattori relativi al numero di fermioni}
which can be expressed in terms of the eigenvalues $h_n$ of the Hamiltonian~$H$ 
%\be
%H\,  f_n(t, {\bf x}) \, = \, h_n\,  f_n(t, {\bf x})
%\ee
by writing
\be
%\frac{\langle  \overline{{\mathcal F}}  | \hat{\sigma} |  \overline{{\mathcal F}}  \rangle} {\langle  \overline{{\mathcal F}} |   \overline{{\mathcal F}}  \rangle}  
\langle \bar{\psi} \psi \rangle_F
=   - \, m \sum_n  \frac{1}{ h_n \sqrt{1 + h_n^2}} %\approx - \, m \sum_{n}  \frac{1}{h_n} 
\label{pre}
\ee
where the largest contribution in the sum comes only from the lowest eigenvalues. 

In order to understand the meaning of this relation, let us consider the direct evaluation $\langle \bar{\psi} \psi \rangle_F$ in the functional integral, that is
%the usual order parameter of chiral invariance, in the functional integral at fixed gauge configuration. This is related to the expectation value of the anti-normally ordered operator $:\hat{\sigma}:$  (see for example~\cite{Lusc}), that is
%\be
%\langle \bar{\psi} \psi \rangle = \frac{\langle  \overline{{\mathcal F}}  | :\hat{\sigma}: |  \overline{{\mathcal F}}  \rangle} {\langle  \overline{{\mathcal F}} |   \overline{{\mathcal F}}  \rangle}  = 2\, \frac{\langle  \overline{{\mathcal F}}  | \hat{\sigma} |  \overline{{\mathcal F}}  \rangle} {\langle  \overline{{\mathcal F}} |   \overline{{\mathcal F}}  \rangle}  \label{psibarpsi}
%\ee
%but,  
(at least in the case $U_0 = \bone$)
\be
\langle \bar{\psi} \psi \rangle_F = - \frac{2}{L_0} \sum_{k=0}^{L_0/2-1}  \sum_{n} \frac{2 m}{h_n^2 + \frac{1}{4}\hat{p}_0^2} 
\ee
where 
\be
\hat{p}_0 = 2 \sin \frac{2 \pi k}{L_0} \,
\ee
is the lattice momentum in the time direction.
%\be
% \frac{1}{V} \int dt \, d {\bf x} \langle \sigma(t, {\bf x}) \rangle_F = - \frac{2m}{V} \sum_{e_n >0}   \frac{1}{m^2 + e_n^2} \,.
%\ee
The sum on $k$ can be exactly performed (see, for example,~\cite[Appendix B]{Pel}), and thus
\be
\langle \bar{\psi} \psi \rangle_F = - \,2\,m\, \sum_{n}   \frac{1}{ h_n \sqrt{1 + h_n^2}} \coth \left[ \frac{L_0}{2} \mathop{\rm arcsinh} h_n \right] \label{last}
\ee
which, at zero temperature, that is in the limit of infinitely large time-direction $L_0$, is in perfect agreement  %with~\reff{psibarpsi} and 
with~\reff{pre}.

But~\reff{last} is the starting point to derive a sort of Banks-Casher relation~\cite{Banks} (see also~\cite[pag.~119]{Kogu}) for static gauge configurations of the form we considered. Indeed, in the limit of vanishing lattice spacing and infinite volume,
\be
\langle \bar{\psi} \psi \rangle_F \approx - \,2\,m\, \sum_{n}   \frac{1}{ h_n} \approx
- \,\frac{m}{\pi}\, \int_0^\infty \frac{dh}{h}\, \rho(h)
\ee
where $\rho(h)$ is the density of energy eigenvalues. Spontaneous symmetry breaking of chiral symmetry is recovered if in the limit of vanishing mass,
\be
- \lim_{m\to 0} \,\frac{m}{\pi}\, \int_0^\infty \frac{dh}{h}\, \rho(h) \neq 0 \, . \label{condition}
\ee
This is the limit which is controlled by the density of eigenvalues near the origin after averaging over the gauge configurations.

In the absence of gauge interaction~\reff{condition} is exactly the Banks-Casher relation.  In presence of the interaction with the gauge fields, in the Banks-Casher relation the spectrum of the Dirac operator is averaged in the full set of gauge configurations (see~\cite{Giusti} for a recent numerical exploration). For our aim we are restricted instead to the energy operator, which does not contain time derivatives, in static gauge-configurations of the form we considered.
In this case the number of dimensions is effectively reduced by one unit and therefore the relation could be more easily checked numerically. Signals that chiral symmetry is spontaneously broken within this restricted ensemble of gauge configurations would be an important check for the effectiveness of our approach.

\section{Quasiparticle  confinement~\label{confinement}}

In this Section we study the propagation of quasiparticles in the vacuum determined in the saddle-point approximation. We remind the reader that this assumes the dominance of chromomagnetic fields, so that the following developments do not apply to the Abelian case.

At the saddle point chromoelectric fields disappear from the pure gauge-field  and mesonic actions, so that the temporal link variables survive only in the action of quasiparticles. Or, in other words, the Gauss constraint still has  to be implemented in the presence of quasiparticles. As we will see, this  can be achieved by exactly performing  the integral on temporal links variables, and leads to color confinement in the quasiparticle sector. 

The quasiparticle action  at the saddle point,  in the $U_0=1$ gauge, if we distinguish the fields at initial and final times, reads 
\begin{multline}
{\overline S}_{qp} =   \frac{1}{2} \left(\alpha^*_0 \alpha_0 + \alpha^*_{\frac{L_0}{s}} \alpha_{\frac{L_0}{s}} - \beta_0 \beta^*_0 -  \beta_{\frac{L_0}{s}} \beta^*_{\frac{L_0}{s}}\right)  + \sum_{t=1}^{L_0/s -1}  \left( \alpha^*_t  \alpha_t - \beta_t \beta^*_t  \right) \\ \, - \sum_{t=0}^{L_0/s -1}  \left( \alpha^*_t \,  e ^{s \mu}\,  {\overline Q}^{-1}  \alpha_{t+1} 
 -\beta_{t+1}  \,  e ^{-s \mu} \, \,  {\overline Q}^{-1}   \beta^*_{t} \right)\, .
\end{multline}
The evaluation of the trace on the Grassmann variables, necessary at finite temperature, induces antiperiodic boundary conditions for the fermion fields
\begin{align}
\alpha^*_{\frac{L_0}{s}} = - \alpha^*_0\, , \qquad & \qquad \alpha_{\frac{L_0}{s}} = - \alpha_0\\
\beta^*_{\frac{L_0}{s}} = - \beta^*_0\, , \qquad & \qquad \beta_{\frac{L_0}{s}} = - \beta_0\, .
\end{align}
The Gauss constraint can be implemented at a given time, say $t=\frac{L_0}{s}$, because it is conserved by the time-evolution. For this purpose we perform a gauge transformation at that time,
\begin{align}
\alpha^*_{\frac{L_0}{s}} \to  \alpha^*_{\frac{L_0}{s}} \, U^\dagger \, , \qquad & \qquad \alpha_{\frac{L_0}{s}} \to  U\, \alpha^*_{\frac{L_0}{s}} \\
\beta^*_{\frac{L_0}{s}} \to U\, \beta^*_{\frac{L_0}{s}}\, , \qquad & \qquad \beta_{\frac{L_0}{s}} \to  \beta_{\frac{L_0}{s}}\, U^\dagger\, .
\end{align}
The integration on $U$ will induce the constraint (for a discussion on the Gauss law in the transfer formalism of lattice gauge theories the interested reader can see~\cite{Smit}; a full discussion for the propagation kernel in the continuum is given in~\cite{RossiTesta}).
The fermion action becomes
\begin{multline}
{\overline S}_{qp} =   \frac{1}{2} \left(\alpha^*_0 \alpha_0 + \alpha^*_{\frac{L_0}{s}} \alpha_{\frac{L_0}{s}} - \beta_0 \beta^*_0 -  \beta_{\frac{L_0}{s}} \beta^*_{\frac{L_0}{s}}\right)  + \sum_{t=1}^{L_0/s -1}  \left( \alpha^*_t  \alpha_t - \beta_t \beta^*_t  \right) \\ - \left( \alpha^*_{\frac{L_0}{s}-1} \,  e ^{s \mu}\,  {\overline Q}^{-1} \, U\, \alpha_{\frac{L_0}{s}} - \beta_{\frac{L_0}{s}} \,U^\dagger \,  e ^{-s \mu} \,  {\overline Q}^{-1} \,  \beta^*_{\frac{L_0}{s} - 1} \right)\, \\
 - \sum_{t=0}^{L_0/s -2}  \left( \alpha^*_t \,  e ^{s \mu}\,  {\overline Q}^{-1}  \alpha_{t+1} 
 -\beta_{t+1}  \,  e ^{-s \mu} \, \,  {\overline Q}^{-1}   \beta^*_{t} \right)
\end{multline}
and using the boundary conditions,
\begin{multline}
{\overline S}_{qp} =   \sum_{t=0}^{L_0/s -1}  \left( \alpha^*_t  \alpha_t - \beta_t \beta^*_t  \right) - \sum_{t=0}^{L_0/s -2}  \left( \alpha^*_t \,  e ^{s \mu}\,  {\overline Q}^{-1}  \alpha_{t+1} 
 -\beta_{t+1}  \,  e ^{-s \mu} \, \,  {\overline Q}^{-1}   \beta^*_{t} \right) \\ 
 + \left( \alpha^*_{\frac{L_0}{s}-1} \,  e ^{s \mu}\,  {\overline Q}^{-1} \, U\, \alpha_0 - \beta_0 \,U^\dagger \,  e ^{-s \mu} \,  {\overline Q}^{-1} \,  \beta^*_{\frac{L_0}{s} - 1} \right)\, .
 \end{multline}
This expression shows that $U$ can be interpreted as the temporal link variable connecting time $\frac{L_0}{s}-1$ with the initial time. The effect of all other temporal link variables has been gauged away. 

We found convenient to perform the change of variables
\begin{align}
 \alpha_t^*  = \gamma_t^* \, , \qquad &  \qquad \alpha_{t} = \begin{cases} -U^{\dagger} \gamma_{\frac{L_0}{s}-1} & \hbox{for \,} t=0\\  \gamma_{t-1} &  \hbox{otherwise} \end{cases} \\
 \beta_t^* =   \delta_t^*\, , \qquad & \qquad  \beta_{t}=  \begin{cases}  -\delta_{\frac{L_0}{s}-1} U & \hbox{for \,} t=0\\  \delta_{t-1} &  \hbox{otherwise} \end{cases}
\end{align}
under which the quasiparticle action transforms into
\begin{multline}
{\overline S}_{qp} =  - \,\gamma_0^*  \,  U_{\frac{L_0}{s}}^{\dagger} \gamma_{\frac{L_0}{s}-1} 
%+  \, \gamma_{\frac{L_0}{s}-1}^* e^{s\mu} {\overline Q}^{-1}   \, \gamma_{\frac{L_0}{s}-1} 
+ \sum_{t=0}^{L_0/s-2 }  \gamma_{t+1}^* \, \gamma_t 
-  \sum_{t=0}^{L_0/s-1} \gamma_t^* e^{s\mu} {\overline Q}^{-1}   \, \gamma_t \\
+\, \delta_{\frac{L_0}{s}-1} U_{\frac{L_0}{s}}  \, \delta_0^* 
%-\, \delta_{\frac{L_0}{s}-1}   e^{-s\mu} {\overline Q}^{-1}   \, \delta_{\frac{L_0}{s}-1}^*
-  \sum_{t=0}^{L_0/s-2 }  \delta_t \, \delta_{t+1}^* 
+  \sum_{t=0}^{L_0/s-1 } \delta_t   e^{-s\mu} {\overline Q}^{-1}   \, \delta_t^*   \,.
\end{multline}
The integral  over $U$ can be performed by using the result obtained in~\cite{Creutz_78} (see also~\cite[pag.~44]{Creu}) about the link integral.
For $SU(N_c)$ matrices
\be
\int  dU  \exp \left[  
\Tr \left(K U^\dagger + J U\right)
%-\gamma_1^*  \,  W_{\frac{L_0}{s} +1}^{\dagger} \gamma_{\frac{L_0}{s}}
%  +\delta_{\frac{L_0}{s}} W_{\frac{L_0}{s}+1}  \, \delta_1^*
 \right]
	      = \exp\left\{ \Tr \left[ K \, \cof \left(\frac { \partial }{ \partial J}\right) \right] \right\} W(J)
\ee
where
\be
 W(J)= \int dU  \exp \left[  \Tr \left(J U\right)  \right]
 =  \sum_{ n=0}^{\infty} c_n (\det J)^n \, .
\ee
The {\em cofactor} of any matrix $A$ is defined by
\be
 (\cof A)_{a b} :=  \frac{1 }{ (N_c-1)!} \,\epsilon_{a a_1 \dots a_{N_c-1}} \epsilon_{b b_1 \dots b_{N_c-1}}
 A_{a_1b_1} \dots A_{a_{N_c-1} b_{N_c-1}}
\ee
so that
\be
\left[A \cdot (\cof A)^T\right]_{ij} = \delta_{ij} \det A \, .
\ee
% and
%\be
% W(J)= \int [ dU_0]  \exp \left[  - J \, \, W_{\frac{L_0}{s} +1}^{\dagger}  \right]_{aa}
% =  \sum_{ n=0}^{\infty} c_n (\det J)^n
%\ee
In order to determine the coefficients $c_n$, we first remark that, if $\partial$ is the matrix with elements $\partial/\partial J_{ij}$,
\be
(\det \partial) W(J) = \int dU  (\det U) \exp \left[  \Tr \left(J U\right)  \right]  = W(J) \label{va}
\ee
because $U\in SU(N_c)$.
But $\det J$ must satisfy the Cayley identity (see~\cite{CSS} for a complete discussion on these identities)
\be
(\det \partial) \, (\det J)^n  \;=\; n(n+1) \cdots (n+N_c-1) \, (\det J)^{n-1} \, \label{Cayley}
\ee
so that the coefficients $c_n$ are determined by~\reff{va} to be
\be
 c_n = \frac{1}{n!}\,\frac{\suf (N_c-1) \suf (n)}{\suf(n + N_c -1)} 
\ee
where $\suf(n)$ is the {\em superfactorial} of $n$, that is
\be
\suf(n) := \prod_{k=1}^n k! = \prod_{k=1}^n k^{n-k+1} \, .
\ee
In our application we have an integral for each spatial site $\bf x$ with sources
\begin{align}
J_{ {\bf x}}^{ a_1, a_2} = &  - \sum_i \left( \gamma_{0, {\bf x}, i }^{a_1}\right)^* \gamma_{\frac{L_0}{s}-1, {\bf x}, i}^{a_2} \\
K_{ {\bf x}}^{ a_1, a_2} =  & \sum_i \delta_{\frac{L_0}{s}-1, {\bf x}, i }^{a_1}\left( \delta_{0, {\bf x}, i}^{a_2}\right)^*     \,.
\end{align}
Then since  in the present case $J_{ {\bf x}}$is nilpotent, with the index of nilpotency $N_J$ equal to the number of  quark intrinsic degrees of freedom, excluding color, the sum over $n$ extends up to $N_J$.
%To ease the notation we define color singlets according to 
%\be
%(\gamma_1 ... \gamma_{N_c}) = \epsilon_{ a_1,...a_{N_c}} 
% \gamma_{1 }^{a_1} ... \gamma_{N_c}^{a_{N_c}} \,.
%\ee
Then
\be
\det \, J_{ {\bf x}} = \frac{ (-1)^{N_c}}{ N_c!} \,
\epsilon_{a_1,\dots,a_{N_c}} \left( \gamma_{0, {\bf x}, i }^{a_1} \dots \gamma_{0, {\bf x}, i }^{a_{N_c}}\right)^*
\epsilon_{b_1,\dots,b_{N_c}} \gamma_{\frac{L_0}{s}-1, {\bf x}, i}^{b_1} \dots \gamma_{\frac{L_0}{s}-1, {\bf x}, i}^{b_{N_c}} 
\ee
which is a linear combination of products of two color singlets at position ${\bf x}$ and times $t=0, \frac{L_0}{s}-1$, respectively.
We see that, at this zero-th order of our perturbative expansion, at variance with $\cal F$-type mesons which already have a finite extension, only pointlike color singlets  of quasiparticles fields can propagate. Indeed, at time $t=1$ there are only color singlets of particles, and at time $t= \frac{L_0}{s}$ only color singlets of antiparticles. 
 Since color is conserved the transfer matrix cannot create colored states.
 
This result can be obtained in a more concrete way by defining the transfer matrix for quasiparticles. To this end we first perform the change of variables
$ \gamma \rightarrow \, e^{-s \mu}\,{\overline Q} \gamma,\, \delta \rightarrow \, e^{s \mu} \, \delta\,{\overline Q}$, and rewrite the quasiparticle action accordingly
\begin{align}
{\mathcal S}_{qp}=&\, -\gamma_0^*  \,  U_{\frac{L_0}{s}}^{\dagger} e^{-s \mu}{\overline Q} \, \gamma_{\frac{L_0}{s}-1}
 +  \sum_{t=0}^{L_0/s-2 }  \gamma_{t+1}^* e^{-s \mu} {\overline Q} \, \gamma_t
	 -   \sum_{t=0}^{L_0/s-1} \gamma_t^*   \, \gamma_t 
	 \nonumber\\  
	 & + \delta_{\frac{L_0}{s}-1} \, U_{\frac{L_0}{s}} e^{s \mu} {\overline Q} \,\delta_0^*   \, 
	-  \sum_{t=0}^{L_0/s-2 } \delta_t  e^{s \mu} {\overline Q} \,\delta_{t+1}^* \,
	 +  \sum_{t=0}^{L_0/s-1 } \delta_t     \, \delta_t^*   \,.
\end{align}
 Then we can write the quasiparticle partition function  in the form
 \be
 Z_{qp}= \int D[\gamma_0^*, \gamma_0, \delta_0, \delta_0] \, \langle U_{\frac{L_0}{ s}}\,\gamma_0 ,\delta_0  \, U_{\frac{L_0}{s}}^{\dagger}
 |  {\mathcal T}_{qp}| \gamma_0 , \delta_0    \rangle
 \ee
 where $\langle U_{\frac{L_0}{ s}}\,\gamma_0 ,\delta_0  \, U_{\frac{L_0}{s}}^{\dagger}
 | $ and $| \gamma_0 , \delta_0    \rangle $ are coherent states and 
 \be
 {\mathcal T}_{qp}= \det \left(  {\overline Q}^{-1}\right) 
 \exp \left(   {\hat \gamma}^{\dagger} \, \ln ( - e^{-s \mu} {\overline Q}) \, {\hat \gamma} + {\hat \delta}^{\dagger} \ln ( - e^{s \mu} {\overline Q})^T \,{\hat \delta}  \right) 
 \ee
 is the quasiparticle transfer matrix.
Integrating over $U$  using the above results we conclude that the Fock space of quasiparticles 
 contains only pointlike color singlets.
 
 In conclusion, ${\mathcal Z}_F $ contains the actions of baryons, antibaryons, and mesons, along with their interactions. The purely mesonic term with
 the smallest number of constituents contains, in the absence of colored mesons of ${\mathcal F}$-type, three quasiparticles and three antiquasiparticles. We notice, however, that different mesonic structures can be constructed in other ways: in terms of diquarks and antidiquarks, as shown in our next work~\cite{Cara2}, or as bound states of ${\mathcal F}$-mesons and quasiparticle-antiquasiparticles.

\section{Further developments  \label{8}}
 
The extraordinary results from lattice QCD push towards the attempts to try to recover
pieces of information about baryonic interactions which are relevant to the phenomenology of atomic nuclei, and which cannot be obtained from phenomenology~\cite{Hats}, such as three-body forces and interactions between nucleons and strange baryons. Our method offers a way to attack these problems. Here we outline a derivation of an action for mesons and nucleons. 
 
In the study of the dynamics of  baryons we perform the nonlinear  change of  variables in the Berezin integrals
 defined in ~\cite{Defr,pal}. For $N_c=3$ it reads
\be
\gamma_{t, {\bf x}, i_1}^{a_1} \, \gamma_{t, {\bf x}, i_2}^{a_2} \, \gamma_{t,{\bf x}, i_3}^{a_3} \sim \epsilon^{ a_1a_2a_3}
h_{i_1, i_2, i_3}^i \, \psi_{t, {\bf x},i} \label{Beretransf}
\ee
where  $\sim$ means equality under the Berezin integral, $h_{i_1, i_2, i_3}^i$ are the baryonic structure functions\cite{Defr} and $\psi_{t, {\bf x},i}$
%\be
%\psi_{t, {\bf x},i} =  \left( \epsilon_{ a_1a_2a_3} \gamma_{t, {\bf x}, i_1}^{a_1} \, \gamma_{t, {\bf x}, i_2}^{a_2} \, \gamma_{t,{\bf x}, i_3}^{a_3}  \right)_i
%\ee
are color singlets from the triplets of Grassmann variables coupled to quantum numbers $i$. The $\psi_{t, {\bf x},i} $ are the new integration variables which are again odd elements of Grassmann algebras.

The expansion is formulated in terms of  mesonic and  baryonic variables only, quarks being altogether eliminated. An application of this change of variables  in a slightly different context can be found in~\cite{Defr}.  We  remark that, within the limitations of validity of the approximation of that calculation, the resulting nucleon action contained a Wilson term as a consequence of the Wilson term for quarks. That effective action, therefore, did not require any additional care to prevent fermion doubling.

 For an illustration we evaluate the contribution quadratic in the baryonic  variables in the present case, neglecting antibaryons and mesons
of non ${\mathcal F}$-type. We must then consider the expression
\be
{\mathcal Z}_F \approx \exp \left(- {\overline S}_{me} \right) \int D[\gamma^*, \gamma]\,  W(J) \exp \left[  \sum_{t=0}^{\frac{L_0}{s}-2}   \gamma_{t+1}^* \gamma_t
- \sum_{t=0}^{\frac{L_0}{s}-1} \gamma_t^* \, {\overline Q}^{-1} \gamma_t \, \right]
\ee
and expand it to third order in both the $\gamma^* $ and $\gamma$
\begin{align}
{\mathcal Z}_F \approx & \exp \left(- {\overline S}_{me} \right) \int D[\gamma^*, \gamma] \,  ( 1 +\det J ) \prod_{t=0}^{\frac{L_0}{s}-1}
\left[ 1 - \frac{1}{3!}  \left( \gamma_t^* \,{\overline Q}^{-1} \gamma_t \right)^3 \right]
\nonumber\\
& \times \prod_{t=0}^{\frac{L_0}{s}-2}\left[ 1+ \frac{1}{3!} \left(\gamma^*_{t+1} \gamma_t\right)^3  \right]\,.
\end{align}
Now we can use the transformations~(\ref{Beretransf}), obtaining  the quadratic approximation in the baryon variables
\be
{\mathcal Z}_F\approx \exp \left(- {\overline S}_{me} \right) \mathcal{J} \int D[\psi^*, \psi] \,  \exp \left(- S_{baryons} \right)
\ee
where  $\mathcal{J}$ is the Jacobian of the transformation~\reff{Beretransf}, explicitly obtained in~\cite{Defr}, and
\be
S_{baryons} = \sum_{t=1}^{\frac{L_0}{s}} \left(  - {\mathcal C}_{ij} \, 
\psi^*_{t+1, {\bf x}, i}\psi_{t, {\bf x}, j}
 + \, \psi^*_{t, {\bf x}, i} \, {\mathcal M}_{{\bf x},i, {\bf y},j} \, \psi_{t, {\bf y}, j} \right) \,.
\ee
The matrices appearing in the above equations are
\begin{eqnarray}
{\mathcal C}_{ij}&=& \sum_{i_1i_2i_3} 36 \, (h^i_{i_1,i_2,i_3})^* h_{i_1,i_2,i_3}^j
\nonumber\\
{\mathcal M}_{{\bf x},i, {\bf y},j} &=& \sum_{i_1i_2i_3,j_1j_2j_3}  \, (h^i_{i_1, i_2, i_3})^* h^j_{j_1,j_2,j_3} \, \epsilon_{a_1a_2a_3}  \epsilon_{b_1b_2b_3}
\nonumber\\
&& \times  \, ({\overline Q}^{-1})^{a_1b_1}_{{\bf x} i_1, {\bf y} j_1}
({\overline Q}^{-1})^{a_2 b_2}_{{\bf x} i_2, {\bf y} j_2}({\overline Q}^{-1})^{a_3b_3}_{{\bf x} i_3, {\bf y} j_3}
\end{eqnarray}
Needless to say, odd powers of Grassmann variables always have nilpotency index
1, and therefore their action cannot be approximated by a nilpotency expansion. 
%We conjecture  that it is essentially  an expansion
%in the inverse number of quark intrinsic degrees of freedom.

It is reasonable to assume that at low energy the important mesons are of ${\mathcal F}$-type. Their interaction with baryons is
\be
{\overline \psi} \left[ \frac{\partial} {\partial {\varphi}(K)} {\mathcal M} \,  \varphi(K)  +
 \frac{\partial} {\partial {\varphi}^*(K)} {\mathcal M} \,  \varphi^*(K) \right] \psi
\ee
so that the mesons-nucleons action is
\be
S_{mn}= S_{me} + S_{baryons} + {\overline \psi} \left[ \frac{\partial} {\partial {\varphi}(K)} {\mathcal M} \,  \varphi(K)  +
 \frac{\partial} {\partial {\varphi}^*(K)} {\mathcal M} \,  \varphi^*(K) \right] \psi 
\ee
where $S_{me}  $ must be expanded in powers of $\varphi^*, \varphi$.

\section{Summary and outlook\label{end}}

In previous works we developed a method of bosonization of  theories with fermions whose low energy excitations are dominated by bosonic modes. We were able to generate composite bosonic fields by transforming  the action of any such theory into
another  exactly equivalent action. The transformed action can be studied in the framework of a nilpotency expansion,  assuming as an asymptotic parameter the index of nilpotency of the composites. The leading approximation is given by saddle-point equations, which determine the properties of the vacuum. In the absence of gauge fields we solved these equations  for both Kogut-Susskind and Wilson fermions.

In the present work we considered the saddle-point equations for the case of gauge theories. This time we found {\it one} exact, gauge covariant solution only for Kogut-Susskind fermions. Such a solution 
 is relevant provided the vacuum is dominated by chromomagnetic fields. From the fermionic point of view this vacuum appears as a condensate of composite bosons which, however, do not have the quantum numbers of chiral fields. We refer to the field of such condensed composites as to a background field. Fluctuations of this background field describe dynamical mesons which we call ${\mathcal F}$-type mesons. They describe chiral mesons, but also other mesons, including colored mesons which should not be observable at zero temperature and baryon density, because of a mechanism which we have not investigated. In such a vacuum live, in addition to  mesonic ${\mathcal F}$-type fields,  fermionic quasiparticles with the quark quantum numbers. Thanks to the background field, quasiparticles do not have any direct coupling with anti-quasiparticles but are  coupled by gauge-field interactions.  

 We then explored some properties of such a vacuum. First we considered the spontaneous breaking of chiral symmetry, and got an expression for the order parameter which is accessible only in the nonperturbative regime, but could be evaluated in a standard Monte Carlo simulation. Second we considered the quasiparticle action. The temporal
 link variables appear in this action in a peculiar way which allowed us to integrate them out exactly. The result is that only pointlike color singlets of the quasiparticle fields can propagate and  therefore have a particle interpretation. They have baryonic or mesonic quantum numbers. Therefore, color is confined in the quasiparticle spectrum. 
 This is a remarkable result in itself, and also because it allows us to introduce such color
 singlets as integration variables, and therefore as fundamental fields,  in the Berezin integral which defines the partition function, using a formalism previously developed. Our  approach, borne to introduce bosonic composites, has also provided a way to introduce fermionic composites, and, therefore, the possibility of formulating QCD in terms of physical fields. In particular we outlined a derivation of a meson-nucleon
 action  from QCD.  
 
 Further investigation of our approach can proceed along several lines. An important issue is the study  of  the action of mesons. A calculation of this kind has already been done for a four-fermion model~\cite{Cara}. Its extension in the presence of gauge fields should allow us to tackle 
 the problem of Goldstone fields, of the chiral anomaly and, of the utmost importance for us, the fate of colored mesons of ${\mathcal F}$-type.
If we could show that in the saddle-point approximation  these mesons are also confined, the nilpotency expansion could be used to describe confinement and dynamics in QCD at the same time and on the same footing.
  
 Related to the above is the  study of the effective action of baryons. This includes interactions of baryons between themselves and with mesons, among which those of ${\mathcal F}$-type should dominate. All these interactions should also give the baryons a finite structure. They are instead pointlike in the saddle-point approximation, at variance with $\cal F$-type mesons of particles-antiparticles, which are already extended objects. 

 Another important issue concerns the theory at finite chemical potential. We  already have some results on this subject, which we will publish
separately~\cite{Cara2}, but we have anticipated them in short form~\cite{Lattice2010}. We  derived an expression of the free energy whose numerical simulation is free of the sign problem. If we make the assumption of the standard theory of color superconductivity  that at sufficiently high values of the chemical potential an expansion with respect to the gauge coupling constant can be justified, we  get results compatible with the standard ones. This finding adds support to the physical relevance of the vacuum we have studied. 

We expect that by increasing the chemical potential and/or the temperature, an increasing number of components of the background field must be set equal to zero, according to a mechanism observed  in the absence of gauge fields~\cite{Palu,Cara1}, until chiral symmetry is recovered and color is deconfined.  If this expectation is verified, the background field will result in a relevant parameter for the definition of QCD phases in the nilpotency expansion. 
 
 Finally, we would like to emphasize that in
comparing the present results with the previous results in the literature, it should be kept in mind that  the contribution of ${\it dynamical}$ fermions is crucial for the vacuum structure in our saddle-point approximation: it changes the QCD vacuum altogether. More explicitly, one cannot compare with calculations in which the energy of the quarks
is evaluated in the vacuum of the pure gauge-field  theory. In particular, instantons are not dominant in the saddle-point approximation, where, instead, magnetic dipole condensation can occur.

\section*{Acknowledgements}

The work of F.~P.~ has been partially 
  supported by EEC under the Contract No.~MRTN-CT-2004-005104.

\appendix

\section{The matrices $M,N$ of the transfer matrix\label{transfer}}

In this Appendix we report the expressions of the matrices $M, N$ appearing in the definition of the transfer matrix for the Kogut-Susskind and  Wilson regularizations. Their common feature is that they  depend only on the spatial link variables.

We write the $\vec{\gamma}$-matrices in terms of the Pauli matrices, adopting 
a convention different from that of L\"uscher~\cite{Lusc}: \be
\gamma_0 = \left(\begin{array}{cc}1 & 0 \\0 & -1\end{array}\right)\, ,
 \qquad \vec{\gamma} = \left(\begin{array}{cc}0 & - i \vec{\sigma} \\ i \vec{\sigma} & 0\end{array}\right)\, ,
 \qquad \gamma_5 = \left(\begin{array}{cc}0 & 1 \\1 & 0 \end{array}\right)\,. \label{gamma}\, .
\ee
Our $\vec{\gamma}$-matrices have, indeed,  opposite sign.

\subsection{ Kogut-Susskind's regularization \label{B}}

Kogut-Susskind fermions in the flavor basis are defined on hypercubes whose sides are  twice the basic lattice spacing. 
While in the text intrinsic quantum numbers and spatial coordinates were comprehensively  represented by one
index $i$, here we distinguish the spinorial index $\alpha=\{1,\ldots,4\}$, the taste index 
$a=\{1,\ldots,4\}$ and the flavour index $i=\{1,...,N_f\}$, while $x=\{t,x_1,\ldots,x_3\}$ is a 4-vector of {\em even} 
integer coordinates ranging in the intervals $[0, L_t-1]$ for the time component and $[0, L_s-1]$ for each of the spatial components.
We distinguish summations over  basic lattice and hypercubes according to
\begin{equation}
\sum_x{}^\prime := 2^d \sum_x \,.
\end{equation}
The projection operators over fermions-antifermion states are
\begin{equation}
P_{\pm} = { \frac{1}{2} } ( \bone \otimes \bone \mp \gamma_0 \gamma_5 \otimes t_5 t_0 ) \, .
\end{equation}
%The relation between the variables $u,v$ and the quark $q$ field is
%\begin{equation}
%P^{(+)}_0 q =\frac{1}{4} u, \qquad P^{(-)}_0 q = \frac{1}{4} v^\dagger \, .
%\end{equation}
The matrix $M=0$ while $N$,  neglecting an irrelevant constant,  is
\be
N =  -2  \Big\{   (m +\sigma)\,  (\gamma_0  \otimes \bone)  
  +{ \sum_{j=1}^3}  (\gamma_0  \gamma_j   \otimes \bone) \left[P^{(-)}_j\nabla_j^{(+)}+ P^{(+)}_j\nabla_j^{(-)}\right]  \vphantom{\sum_{j=1}^3} \Big\}
  \label{etacoupling}
\ee
where $\sigma$ is a scalar field and 
\begin{eqnarray}
 \nabla_j^{(+)} & = & \frac{1}{2} \left( U_j \,T^{(+)}_j  - \bone \right) \\
 \nabla_j^{(-)} & = &  \frac{1}{2} \left( \bone - T^{(-)}_j  U_j^{\dagger}\right)
 \end{eqnarray}
 are the lattice covariant derivative and~\footnote{There is a misprint in formula~\cite[(A.6)]{Cara1}.}
\begin{equation}
P_j^{(\pm)} = { \frac{1}{2} } ( \bone \otimes \bone \pm \gamma_j \gamma_5 \otimes t_5 t_j ) \, .
\end{equation}
The lattice hamiltonian  $H$ is related to $N$  by
\begin{equation}
 H^2 = \frac{1}{4}  N^{\dagger} N 
 %=  m ^2 -  \Delta 
 \, .
 \label{pippo}
\end{equation}
%where 
%\begin{equation}
%\Delta = \frac{1}{4}   \sum_{j=1}^3 \left(  U_j \,T_j^{(+)}+ T_j^{(-)} U_j \,^\dagger -2 \cdot \bone\right) \,.   \label{laplascian}
%\end{equation}
Then~\cite{Cara}
\begin{equation}\label{GSbosoncompo}
A =  (2\, H)^{-1} \left (H+\sqrt{1+H^2} \right)
\end{equation}
and using this expression, we derive 
\begin{align*}
\overline{\cal H} \, = & \,  e^{s\mu}\, H\, \left( \sqrt{1 + H^2} - H \right) \\
\giro{\overline{\cal H}} \,  = & \, e^{-s\mu}\, H\, \left( \sqrt{1 + H^2} - H \right)
\end{align*}
so that in the formal continuum limit
\be
\overline{\cal H} \, \approx  \, \giro{\overline{\cal H}} \, \approx \, H %=  \sqrt{ m^2 - \Delta } 
\ee
both approach the same value.

\subsection{ Wilson's regularization }

The projection operators over fermions-antifermions are
\begin{equation}
P_{\pm} = { \frac{1}{2} } ( \bone \pm \gamma_0 ) \,.
\end{equation}
%The relations between the quark field $q$ and its upper and lower components $u,v$  are
%\begin{equation}
%P^{(+)}_0 q = B^{-\frac{1}{2} }\, u, \,\,\, P^{(-)}_0 q = B^{-\frac{1}{2} } v^\dagger\,,
%\end{equation}

The matrices $M, N$ are
\begin{eqnarray}
M&=&  { \frac{1}{2} } \ln \left( \frac{B}{2K} \right)\\
N &= & 2 K \, B^{- \frac{1}{2} }\,c \, B^{-\frac{1}{2} } \,,
\end{eqnarray}
where
\begin{equation}
B = 1 -  K \sum_{j=1}^3 \left (U_j T^{(+)}_j + T^{(-)}_j U^\dagger_j  \right) 
 \end{equation}
$K$ is the hopping parameter and
\begin{equation}
c = { \frac{1}{2} }  \sum_{j=1}^3 i \left ( U_j  \, T^{(+)}_j - \, T^{(-)}_j U^\dagger_j  \right)\,
\sigma_j \,.
\end{equation}


\begin{thebibliography}{11}


\bibitem{Wilson}
K.~G.~Wilson,
{\em Confinement of quarks},
Phys.\ Rev.\ D {\bf 10}  (1974) 2445.

\bibitem{MM}
I.~Montvay and G. M\"unster,
{\em Quantum Fields on a Lattice}, 
Cambridge Monographs on Mathematical Physics, Cambridge University Press, 1994.

\bibitem{Palu05}
F.~Palumbo,
{\em Boson dominance in nuclei},
Phys.\ Rev.\  C {\bf 72} (2005) 014303.
%%CITATION = PHRVA,C72,014303;%%

\bibitem{Cara}
S.~Caracciolo, V.~Laliena and F.~Palumbo,
{\em Composite boson dominance in relativistic field theories},
JHEP {\bf 0702} (2007) 034 [arXiv:hep-lat/0611012].
%%CITATION = JHEPA,0702,034;%%

\bibitem{Palu}
F.~Palumbo, 
{\em A semi-variational approach to QCD at finite temperature and baryon density},
Phys.\ Rev.\ D  {\bf  78} (2008) 0145514   [arXiv:hep-lat/0702001]. 
Please note that, in this paper, the chemical potential for Kogut-Susskind fermions must be multiplied by a factor of 2.
%%CITATION = PHRVA,D78,014514;%%

%\bibitem{Arak}
%H.~Araki,  
%{\em On the diagonalization of a bilinear Hamiltonian by a Bogoliubov transformation}, 
%Publ. RIMS, Kyoto Univ. Ser. A Vol. 4 (1968);
%W.~A.~Bardeen, 
%{\em Schr\"odinger approach to ground state wavefunction},  
%in Proc. Int. Workshop on Variational Calculus in Quantum Field Theory, Wangerooge, West 
%Germany, Sept. 1-4, Singapore World ScientiÞc (1987); 
%W.~G.~Unruth, 
%{\em Notes on black hole evaporation}, 
%Phys.\ Rev. \ D {\bf 14} (1976) 870;
%%%CITATION = PHRVA,D14,870;%%
%R.~M.~Wald, 
%{\it Quantum filed theory in curved spacetime and black holes thermodunamics}, 
%University of Chicago Press (1994);
\bibitem{Misha}
B.~Chatterjee, H.~Mishra and A.~Mishra, 
{\em BCS-BEC crossover and phase structure of relativistic systems: A variational approach},
Phys.\ Rev.\  D {\bf 79} (2009) 014003.

\bibitem{Marchetti}
G.~Jona-Lasinio and F.~M.~Marchetti,
{\em On the pairing structure of the vacuum induced by a magnetic field in 2+1-dimensional Dirac field theory},
Phys.\ Lett.\  B {\bf 459} (1999) 208
[arXiv:hep-th/9906014].
%%CITATION = PHLTA,B459,208;%%


\bibitem{Cara1}
S.~Caracciolo,  F.~Palumbo and G.~Viola,
{\em Bogoliubov transformations and fermion condensates in lattice field theories},
Annals Phys.\  {\bf 324} (2009) 584 [arXiv:0808.1110 [hep-lat]].
%%CITATION = APNYA,324,584;%%

\bibitem{Lattice2010}
S.~Caracciolo and F.~Palumbo, 
{\em Absence of sign problem in the (saddle point approximation of the) nilpotency expansion of QCD at finite chemical potential},
PoS (Lattice 2010) 210 [arXiv:1011.0581], 
Proceedings of The XXVIII International Symposium on Lattice Field Theory, June 14-19, 2010, Villasimius, Italy. 
%%CITATION = 1011.0581;%%

\bibitem{Savvidy}
G.~K.~Savvidy,
{\em Infrared Instability of the Vacuum State of Gauge Theories and Asymptotic Freedom},
Phys.\ Lett.\  {\bf B71 } (1977)  133.

\bibitem{Nielsen}
N.~K.~Nielsen and P.~Olesen,
{\em An Unstable Yang-Mills Field Mode},
Nucl.\ Phys.\  {\bf B144 } (1978)  376.

\bibitem{Nambu}
Y.~Nambu,
{\em Strings, Monopoles and Gauge Fields},
Phys.\ Rev.\  {\bf D10 } (1974)  4262.

\bibitem{Pari}
G.~Parisi, 
{\em Quark imprisonment and vacuum repulsion}, 
Phys.\ Rev.\ {\bf D} 11 (1975) 970.

\bibitem{Mande}
S.~Mandelstam, 
{\em Charge-monopole duality and the phases of non-Abelian gauge theories},
Phys.\ Rep.\ {\bf 23} (1976) 245.

\bibitem{Thooft} 
G.~'t Hooft, 
{\em Gauge theories for strong interactions}, in New Phenomena in Subnuclear Physics, edited by A. Zichichi (Plenum Press, New York, 1977.

\bibitem{Suganuma}
H.~Suganuma, S.~Sasaki and H.~Toki,
{\em Color confinement, quark pair creation and dynamical chiral symmetry breaking in the dual Ginzburg-Landau theory},
Nucl.\ Phys.\  {\bf B435 } (1995)  207-240  [arXiv:hep-ph/9312350].

\bibitem{Kogu}
J. B. Kogut and M. A. Stephanov, 
{\em The Phases of Quantum Chromodynamics: From Confinement to Extreme Environments}, Cambridge University Press, 2004.


\bibitem{Poli}
P.~V.~Buividovich, M.~N.~Chernodub, D.~E.~Kharzeev, T.~Kalaydzhyan,  E.~V.~Luschevskaya and M.~I.~Polikarpov, 
{\em  Magnetic-Field-Induced insulator-conductor transition in SU(2) quenched lattice gauge theory}
Phys.\ Rev.\ Lett.\ {\bf 105} (2010) 132001 [arXiv:1003.2180].

\bibitem{Delia}
M.~D'Elia and F.~Negro,
{\em Chiral properties of strong interactions in a magnetic background},
[arXiv:1103.2080].

\bibitem{Lusc}
M.~L\"uscher,
{\em Construction Of A Selfadjoint, Strictly Positive Transfer Matrix For Euclidean Lattice Gauge Theories},
Commun.\ Math.\ Phys.\  {\bf 54} (1977) 283.
%%CITATION = CMPHA,54,283;%%

\bibitem{Palu-KS}
F.~Palumbo,
{\em The transfer matrix with Kogut-Susskind fermions},
Phys.\ Rev.\  D {\bf 66} (2002) 077503
[Erratum-ibid.\  D {\bf 73} (2006) 119902]
[arXiv:hep-lat/0208005].
%%CITATION = PHRVA,D66,077503;%%

\bibitem{Cara2}
S.~Caracciolo and F.~Palumbo, to be published.
 
\bibitem{Houa}
J.-C.~Houard and B.~Jouvet, 
{\em Etude d'un mod\`{e}le de champ \`{a}  constante de renormalisation nulle},
Il Nuovo Cimento {\bf 18} (1960) 466.

\bibitem{Salam} 
A.~Salam, 
{\em Lagrangian theory of composite particles},
Il Nuovo Cimento {\bf 25} (1962) 224.

\bibitem{Weinberg}
S.~Weinberg, {\em The Quantum Theory of Fields}, Cambridge University Press, 1995.
 
\bibitem{Vers}
D.~Verstegen, 
{\em Symmetry properties of fermionic bilinears in lattice QCD}, 
Nucl.\ Phys.\ B {\bf 249} (1985) 685.
%%CITATION = NUPHA,B249,685;%%

\bibitem{Golt}
M.~Golterman, 
{\em Staggered mesons}, 
Nucl.\ Phys.\ B {\bf 273} (1986) 663.
%%CITATION = NUPHA,B273,663;%%

\bibitem{Pel}
S.~Caracciolo and A.~Pelissetto,
{\em Corrections to finite-size scaling in the lattice N-vector model for  N = $\infty$},
Phys.\ Rev.\  D {\bf 58} (1998) 105007
[arXiv:hep-lat/9804001].
%%CITATION = PHRVA,D58,105007;%%

\bibitem{Banks}
T.~Banks and A.~Casher,
{\em Chiral Symmetry Breaking In Confining Theories},
Nucl.\ Phys.\  B {\bf 169} (1980) 103.
%%CITATION = NUPHA,B169,103;%%

\bibitem{Giusti}
L.~Giusti and M.~L\"{u}scher,
{\em Chiral symmetry breaking and the Banks--Casher relation in lattice QCD with Wilson quarks},
JHEP {\bf 0903} (2009) 013
[arXiv:0812.3638].
%%CITATION = JHEPA,0903,013;%%

\bibitem{Smit}
J.~Smit,
{\em Introduction to Quantum Fields on a Lattice},
Cambridge Lecture Notes in  Physics 15, Cambridge University Press,  2002.

\bibitem{RossiTesta}
G.~C.~Rossi and M.~Testa,
{\em The Structure of Yang-Mills Theories in the Temporal Gauge (I). General formulation},
Nucl.\ Phys.\ B {\bf 163} (1980) 109.
%%CITATION = NUPHA,B163,109;%%

\bibitem{Creutz_78}  
M.~Creutz, 
{\em On invariant integration over $SU(N)$},
J.\ Math.\ Phys.\ {\bf 19}, (1978) 2043.

\bibitem{Creu}
M.~Creutz, 
{\em Quarks, gluons and lattices}, 
Cambridge Monographs on Mathematical Physics, Cambridge University Press, 1983.

\bibitem{CSS}
S.~Caracciolo, A.~D.~Sokal and A.~Sportiello,
{\em Combinatorial proofs of Cayley-type identities for derivatives of determinants and pfaffians}, [arXiv:1105.6270].
%%CITATION = 1105.6270;%%

\bibitem{Hats}
T.~Hatsuda, 
{\em Nuclear Physics on the lattice},
PoS (Lattice 2010) 008, 
Proceedings of The XXVIII International Symposium on Lattice Field Theory, June 14-19, 2010, Villasimius, Italy. 

\bibitem{Defr}
G.~De Franceschi and F.~Palumbo, 
{\em Composite operators as integration variables in Berezin integrals}, 
Modern Phys.\ Lett.\ A {\bf 10} (1995) 901.

\bibitem{pal}
F.~Palumbo, 
{\em Quark composites approach to QCD: The nucleon-pion system}, 
Phys.\ Rev.\ D {\bf 60} (1999) 074009.




\end{thebibliography}
\end{document}